\documentclass[twocolumn,aps]{revtex4}

\usepackage{amsfonts}
\usepackage{amsmath}
\usepackage{amssymb}
\usepackage{graphicx}
\usepackage{dcolumn}
\usepackage{bm}
\usepackage{color}

\begin{document}

\title{Enhancement of the Curie temperature in small particles of weak itinerant ferromagnets}
\author{L.~Peters, M. I. Katsnelson, and A.~Kirilyuk}
\affiliation{Radboud University Nijmegen, Institute for Molecules and Materials, NL-6525 AJ Nijmegen, The Netherlands}
\date{\today}

\begin{abstract}

Self consistent renormalization theory of itinerant ferromagnets
is used to calculate the Curie temperature of clusters down to
approximately 100 atoms in size. In these clusters the electrons
responsible for the magnetic properties are assumed to be (weakly)
itinerant. It is shown that the Curie temperature can be larger
than in the bulk. The effect originates from the phenomenon of
level repulsion in chaotic quantum systems, which suppresses spin
fluctuations. Since the latter destroy the magnetic order the
resulting Curie temperature increases, contrary to expectations of
the na\"\i ve Stoner picture. The calculations are done assuming
that the energy levels of the cluster are described by the
Gaussian Orthogonal Ensemble of random matrix theory.

\end{abstract}

\maketitle

\section{Introduction}
Nowadays, nanotechnology is one of the most promising research
areas. By using atoms and molecules as building blocks or by down
scaling macroscopic objects, a large variety of nano-scale objects
can be created. These nano-scale objects possess interesting
physical properties substantially differing from the bulk
\cite{Khanna}. It is size quantization of energy levels that could
cause, sometimes, a dramatic difference of optical, electronic or
chemical properties compared to the bulk. Applications in material
science, health care, etc. are offered for these nano-objects
\cite{Nn}.

Studying clusters containing tens to tens of thousands of atoms is
a subject of special interest \cite{Sugano}. Such small particles
demonstrate unusual mechanical, optical, and magnetic properties
and chemical activity. The latter is crucial, for example,  for
their use in catalysis processes. However, the precise mechanisms
behind the physical and chemical properties of the clusters are
not yet fully understood \cite{Khanna}. For new applications and
improvements of existing ones it is extremely important to
understand these physical principles.

The study of magnetic properties is one of the \mbox{methods} to
gather information about the cluster's behavior. Namely, the
magnetic properties are sensitive to atomic and electronic
structure, quantum size effects, surface to volume ratio, and
symmetry. Several experiments demonstrate unusual magnetism of
metallic clusters. For example, it was found that the clusters of
3d transition metals iron, cobalt and nickel, show enhanced
ferromagnetism in comparison with the bulk
\cite{Liu,Bucher,Billas}. Further, rhodium clusters appeared to
become ferromagnetic while in the bulk rhodium is paramagnet
\cite{Cox}. For manganese, a change of magnetic order from ferro-
to antiferromagnetism occurs during the transition from cluster to
the bulk regime \cite{Nickelbein}.

In general it is a complicated matter to understand the physical
principles behind the magnetic properties of metallic clusters.
Within the picture of localized magnetic moments described by the
Heisenberg model one could expect that the smaller the cluster
size and, thus, the larger the ratio surface to bulk, the weaker
the magnetism is, due to a lower coordination of surface atoms.
For an itinerant electron system the influence of downscaling is
not immediately clear. For certain small cluster sizes so called
shell effects give rise to an increased density of states near the
Fermi level \cite{Kresin} and thus, due to the Stoner criterion
\cite{Moriya}, an increased tendency towards ferromagnetism.
Besides an influence on the density of states, the size could also
have an effect on the thermal spin fluctuations responsible for
the reduction of magnetic order, whose role is crucial for the
understanding of the magnetic properties at finite temperatures
\cite{Moriya}.

Note that in ultra thin films finite size effects also cause interesting
physical features such as changes in the Curie temperature
\cite{Aguil,Irkk}. For example in Gd ultra thin films an increased
surface Curie temperature is observed compared to the bulk
\cite{Aguil}.

The theory of itinerant electron magnetism is, in general, very
complicated and still controversial (for a review of some modern
approaches, see Refs. 13, 14 and references therein). For a
special case of weak itinerant ferromagnets close enough to the
point of the Stoner instability a very successful self consistent
renormalization (SCR) theory has been developed
\cite{Moriya,Dzya}. This theory clarifies the role of spin
fluctuations in finite-temperature magnetism and allows us to
calculate the Curie temperature. Here we use this theory to study
the size effects on the Curie temperature for metallic clusters,
down to approximately 100 atoms in size. It appears that
suppression of the spin fluctuations leads to an increase of the
Curie temperature when going to smaller cluster sizes.

\section{The self consistent renormalization theory of weak itinerant ferromagnetism}
For a convenience of readers, we present in this section a brief
overview of the SCR theory, following Ref. \onlinecite{Moriya}. We
deal with the simplest model of itinerant electron magnetism, that
is, the Hubbard model with the Hamiltonian
\begin{equation}
\begin{split} \label{Hubbard}
&H=\sum_{j,l}\sum_{\sigma}t_{jl}c_{j\sigma}^{\dagger}c_{l\sigma}+U\sum_jn_{j\uparrow}n_{j\downarrow} \\
&=\sum_{k}\sum_{\sigma}\epsilon(k)a_{k\sigma}^{\dagger}a_{k\sigma}+I\sum_q\sum_k\sum_{k'}a_{k+q\uparrow}^{\dagger}a_{k'-q\downarrow}^{\dagger}a_{k'\downarrow}a_{k\uparrow}. \end{split}
\end{equation}
Here $t_{jl}$ is the transfer integral between the Wannier
orbitals at sites $j$ and $l$, $\epsilon(k)$ is its Fourier
transform, that is, the band energy dependent of the wave vector
$k$, $a_{k\sigma}$ is the annihilation operator for the Bloch
state, $c_{l\sigma}$ is the annihilation operator for the Wannier
state, $n_{j\uparrow}$ is the occupation number operator for the
Wannier state at site $j$ with spin-up, $U$ is the Coulomb
interaction energy \mbox{between} two electrons on the same site
and $I=\frac{U}{N_0}$ with $N_0$ the total number of atoms. Thus,
the first term of the Hubbard model is the hopping term or
tight-binding model and the second term is the on-site Coulomb
interaction term.

The simplest way to approximate the interaction term is by using
the Hartree-Fock approximation, which leads to the familiar Stoner
theory:
\begin{equation}
n_{j\uparrow}n_{j\downarrow} \stackrel{HFA}{\longrightarrow}n_{j\uparrow}\langle n_{j\downarrow}\rangle+n_{j\downarrow}\langle n_{j\uparrow}\rangle-\langle n_{j\uparrow}\rangle\langle n_{j\downarrow}\rangle. \label{approx}
\end{equation}
Here the $\langle \ldots \rangle$ denotes the statistical average
of the quantity inside the brackets. For the case of a
ferromagnet, the statistical average of the occupation number is
supposed to be site-independent: $\langle
n_{j\uparrow}\rangle=\langle n_\uparrow \rangle=n_\uparrow$. Using
the approximation of \eqref{approx} for Eq. \eqref{Hubbard} and by
rewriting this expression, leads to

\begin{equation}
H=\sum_k\sum_{\sigma}E_{k\sigma}a_{k\sigma}^{\dagger}a_{k\sigma}+\frac{1}{4}IN^2-IM^2. \label{HFA}
\end{equation}
Here $N$ is the total number of electrons, $M$ is the
magnetization and $E_{k\sigma}$ is the single electron energy
given by
\begin{equation}\label{ED}
\begin{split}
&E_{k\sigma}=\epsilon_k+\sigma\Delta, \\
&\Delta=IM+\mu_B\tilde{H}
\end{split}
\end{equation}
with $\sigma$ equal to $1$ for spin-up and to $-1$ for spin down,
$\mu_B$ the Bohr magneton and $\tilde{H}$ an external magnetic
field. Further, $2\Delta$ is the band splitting between the
spin-up and spin-down bands. This ferromagnetic state is stable at
temperature $T=0$ if $I\rho(E_F)>1$ (Stoner criterion) with
$\rho(E_F)$ the density of states (per whole system per spin) at
the Fermi level.

From the results of Eqs. \eqref{HFA} and \eqref{ED} it is possible
to derive the thermodynamic properties of the system in the Stoner
theory \cite{Moriya}. The Curie temperature, which is of main
interest here, can be calculated from the divergence of the static
magnetic susceptibility of the system, which within the Stoner
theory is given by
\begin{equation}
\chi(T)=\frac{\chi_0(T)}{1-I\chi_0(T)}. \label{XS}
\end{equation}
Here $\chi_0$ is the static magnetic susceptibility of a
non-interacting system (i.e. with $I=0$). For the bulk $\chi_0$ is
given by
\begin{equation}
\chi_0(T)=\rho(E_F)\left[1-\frac{\pi^2}{6}RT^2+\ldots\right]. \label{NIS}
\end{equation}
with
$R=\left(\frac{\rho'}{\rho}\right)^2-\left(\frac{\rho''}{\rho}\right)\Big\vert_{E=E_F}$
and the accent corresponds to the derivative to $E$.

Thus, from Eq. \eqref{XS} and the divergence of the static
magnetic susceptibility the Curie temperature $T_C$ follows from
\begin{equation}
1-I\chi_0(T_C)=0. \label{CE}
\end{equation}
For the case of weak itinerant ferromagnets when
\begin{equation}
0<I\rho(E_F)-1\ll1. \label{DefW}
\end{equation}
the Stoner Curie temperature \eqref{CE} is estimated as
\begin{equation}
T_C^S\propto E_F\sqrt{I\rho(E_F)-1}. \label{TCS}
\end{equation}
It is well known that Curie temperatures predicted from this
equation are too high compared to the experimental data
\cite{Moriya,Lichtenstein}. The reason is a neglect of spin
fluctuations that allow magnetic excitations at lower energy: it
is easier to rotate the spins than to change their length like in
the Stoner theory.

The SCR theory that takes spin fluctuations into account is
formulated in the following way \cite{Moriya} (an alternative
approach based on diagram technique has been developed in Ref.
13). The transverse dynamic magnetic susceptibility
$\chi^{-+}(q,\omega)$ can be formally represented as
\begin{equation}
\chi^{-+}(q,\omega)=\frac{\chi_0^{-+}(q,\omega)}{1-I\chi_0^{-+}(q,\omega)+\lambda_{MI}(q,\omega)}.\label{EDS}
\end{equation}
Here $\chi_0^{-+}(q,\omega)$ is the transverse dynamic magnetic
susceptibility for non-interacting electrons with spin split
bands, $q$ is the wave-vector and $\omega$ the frequency. In
general the problem is to find $\lambda_{MI}(q,\omega)$ of Eq.
\eqref{EDS} to make the expression for the dynamic susceptibility
exact. For this purpose the total free energy $F$ of a system is
expressed in terms of the exact transverse dynamic susceptibility
using the Hellmann-Feynman theorem:
\begin{equation}
\begin{split}
&F=F_0+\Delta F\\
&=F_0-\frac{1}{2\pi}\int_{-\infty}^\infty d\omega\coth\left(\frac{\omega}{2T}\right)\sum_q\int_0^IdI\mathrm{Im}\left[\chi^{-+}(q,\omega)\right]. \label{F}
\end{split}
\end{equation}
Here $F_0$ is the free energy of a non-interacting system. Then,
one can use the self consistency condition to find
$\lambda_{MI}(q,\omega)$. This means that the static magnetic
susceptibility calculated from the free energy of Eq. \eqref{F}
(via $\frac{\partial^2F}{\partial^2M}=\frac{1}{\chi}$) is equal to
the static long wave-length limit of the exact transverse dynamic
magnetic susceptibility of Eq. \eqref{EDS}.

Of course, this condition is not enough to find the whole
\textit{function} $\lambda$. However, for the weak itinerant
systems where the condition \eqref{DefW} is fulfilled one can
assume that $\lambda$ can be considered in Eq. \eqref{F} as a
\textit{number} independent on $M$, $I$, $\omega$ and $q$; a
formal justification has been given in Ref. 15. As a result,
$\lambda$ is given by the expression \cite{Moriya}
\begin{equation} \label{L1}
\begin{split}
&\lambda(T,d)=\frac{1}{2\pi}\int_{-\infty}^{\infty}d\omega\coth\left(\frac{\omega}{2T}\right)\mathrm{Im}\left\{G(\omega,d)\right\}, \\
&G(\omega,d)=-\alpha\chi_0\sum_q\left[\vphantom{\frac{(d+1)\left(\frac{\partial f_M}{\partial M}\right)^2}{(d+1-f_M)^2}}\frac{f_M\left(\frac{\partial^2f_M}{\partial^2M}\right)}{d+1-f_M}\right. \\ &\qquad\qquad\qquad\qquad\qquad \left. +\frac{(d+1)\left(\frac{\partial f_M}{\partial M}\right)^2}{(d+1-f_M)^2}\right]_{M=0}.
\end{split}
\end{equation}
Here $\alpha=I\chi_0$, $f_M=\chi_0^{-+}(q,\omega)/\chi_0$, $d$ is
defined as $\chi_0/\chi\equiv\alpha d=1-\alpha+\lambda(T,d)$ and
$\chi_0\equiv\chi_0^{-+}(0,0)$. At the Curie temperature $d=0$ and
$M=0$, which makes it natural to expand $f_M$ in terms of small
$\omega$ and $q$ ($\omega\ll E_F$ and $q\ll k_F$ respectively with
$k_F$ the Fermi wave-vector) and to approximate the nominators of
Eq. \eqref{L1} by their static and long wavelength limit. Thus, at
the Curie temperature $f_M$ can be approximated for the bulk by

\begin{equation}
f_0=1-Aq^2+iC\frac{\omega}{q},
\label{fbulk}
\end{equation}
where $A$ and $C$ are constants (depending on the shape of the
Fermi surface), and the subscript zero refers to $M=0$
\cite{Moriya}.

By using the condition of divergence of the static magnetic
susceptibility, the Curie temperature for a weak itinerant system
is given by the equation
\begin{equation}
1-I\chi_0(T_C)+\lambda(T_C,0)=0. \label{SCRTC}
\end{equation}
Here, it can be noticed that it is actually $\lambda$ which takes
the influence of the spin density fluctuations on the Curie
temperature into account. For further calculations it is
convenient to separate $\lambda$ in the following two parts:
\begin{equation} \label{Lsplit}
\begin{split}
&\lambda(T,0)=\lambda_0+\lambda_1, \\
&\lambda_0=\frac{1}{2\pi}\int_{-\infty}^\infty d\omega\, \text{sign}\{\omega\}\text{Im}\left\{G(\omega,0)\right\},\\
&\lambda_1=\frac{1}{2\pi}\int_{-\infty}^\infty d\omega\, \text{sign}\{\omega\} \frac{2}{e^{\frac{|\omega|}{T}-1}}\text{Im}\left\{G(\omega,0)\right\}.
\end{split}
\end{equation}
The function $\lambda_0$ is the temperature independent part,
which merely gives rise to a shift of the Stoner criterion at $T
=0$. It can be simply considered as a renormalization of the
Stoner parameter $I$.  Strictly speaking, $\lambda_0$ does depend
on temperature via $\chi_0$, but its temperature dependence can be
neglected compared to that of $\lambda_1$. Namely, one can show
that $\lambda_1\propto T^{4/3}$ compared to the $T^2$ dependence
of $\chi_0$ \cite{Moriya}. Important is that due to this
temperature dependence of $\lambda_1$ the Curie temperature is
effectively lowered compared to Stoner theory:
\begin{equation}
T_C\propto E_F\left[I\rho(E_F)-1\right]^{3/4}\approx\left(T_C^S\right)^{3/2}/E_F^{1/2}. \label{SCR-TC}
\end{equation}
In other words, the dynamics of the spin density fluctuations is
crucial for the correct description of the Curie temperature.

\section{Size dependent energy level distribution}

The SCR theory described above will be used further to calculate
the Curie temperature of metallic clusters as a function of their
size. Note that the transverse dynamic magnetic susceptibility of
a non-interacting system is in fact the only input required for
this calculation. For clusters this quantity substantially differs
compared to the bulk due to the differing energy level spectrum.
The exact calculation of the energy spectrum of a cluster with a
given size is actually impossible due to certain randomness of its
shape. To overcome this problem, the random matrix theory
\cite{Peerenboom,Halperin,Stockmann,Dyson,Bohr} is typically used.

In principle it is possible to calculate the well defined energy
levels of an individual small particle. For the case of a perfect
metallic sphere this is actually quite obvious. One obtains an
energy level spectrum consisting of highly degenerate energy
levels, where the separation between the {\it groups} of the
energy levels is proportional to $\frac{1}{L^2}$ with $L$ the
diameter of the sphere and the high degeneracy is due to the high
geometric symmetry. It is worthwhile to stress, however, that
although the nonzero level splitting is proportional to
$\frac{1}{L^2}$, the {\it average} energy level separation around
the Fermi level follows the well known $\frac{1}{V}$ $\propto$
$\frac{1}{L^3}$ proportionality (with $V$ the volume).

Obviously the situation of the perfect sphere is very special.
However, there are examples of clusters with high geometric
symmetry \cite{Kresin}. These are quasi-spherical clusters with so
called ``magic'' numbers $\bar{N}=\bar{N}_m$, where $\bar{N}$ is
the number of atoms. For example $\bar{N}_m=13$, $55$ and $147$
for the Mackay icosahedron, where each ``magic'' number
corresponds to the right number of atoms so that a spherically
shaped cluster can be formed by packing the Mackay icosahedrons in
the proper way. Characteristic for the energy level spectrum of
these ``magic'' clusters is the shell structure. These are highly
degenerate or close groups of energy levels, which causes the
energy level separation near the Fermi level to be rather small.
The effects of this shell structure on the electronic pairing in
superconductors has been discussed in Ref. \onlinecite{Kresin}.
For the Curie temperature of the magic clusters the smaller energy
level separation around the Fermi level could have important
consequences, too.

Here a generic case will be considered only, meaning that the
situation of the highly geometrically symmetric clusters is
excluded. It is assumed that there are uncontrollable atomic
surface irregularities, which are sufficient to split apart this
large degeneracy of the energy levels. Further, it is assumed that
the clusters are large enough to satisfy the proportionality
\mbox{$\delta=1/\rho(E_F)\propto1/V$, where $\delta$} is the
average energy level spacing around the Fermi level and $V$ is the
volume of the cluster \cite{Peerenboom,Halperin,Stockmann}. Thus,
considering an ensemble of clusters of the same size, they will
differ in their energy level spectrum due to the uncontrollable
surface, but have the same average energy level spacing around the
Fermi level.

Gor'kov and Eliashberg \cite{Gor} were the first who recognized
that this situation is similar to the interpretation of nuclear
energy level spectra discussed by Wigner and Dyson \cite{Bohr}.
The idea was to circumvent the unknown and complex interactions
between the nucleons by using a statistical description leading to
an energy level distribution. To be more specific, it was assumed
that the eigenvalues of a random matrix could be taken as a model
for the energy levels of a complex nuclear system. This means that
an ensemble of possible nuclear systems corresponds to an ensemble
of random matrices. Important to remark is that the randomness of
each matrix is restricted, because they must possess certain
transformation properties imposed by the symmetries that each
individual Hamiltonian is supposed to have in common. Then,
depending on the imposed symmetry properties different energy
level distributions can be derived \cite{Bohr}.

In the same manner as described above the uncontrollable surface
irregularities can be interpreted as giving rise to a random
matrix treatment. For metallic clusters the transformation
properties of the random matrix are determined by the magnitude of
the spin-orbit coupling $H_{so}$ and the magnitude of the Zeeman
splitting in an external magnetic field $2\mu_B\tilde{H}$ compared
to $\delta$. For example for small $H_{so}$ and small
$2\mu_B\tilde{H}$ compared to $\delta$ the matrix must have
respectively rotational and time-reversal invariance
\cite{Peerenboom,Halperin,Stockmann,Dyson,Bohr}. This example
corresponds to the so called Orthogonal ensemble, which is used in
this work. The other possible ensembles are given in Table
\ref{Ensembles}.

\begingroup
\squeezetable
\begin{table}[!h]
\caption{overview of different Hamiltonian symmetries relevant for energy level distribution}
\begin{tabular}{|l|l|l|}
\hline
\textbf{Probability distribution} & \textbf{Magnetic field} & \textbf{Spin-orbit coupling} \\
\hline
Poisson & Large & Small \\
\hline
Orthogonal & Small & Small \\
& Small & Large (even particles) \\
\hline
Unitary & Large & Large \\
\hline
Sympletic & Small & Large (odd particles) \\
\hline
\end{tabular}
\label{Ensembles}
\end{table}
\endgroup
The Poisson ensemble is typical for systems with a regular
classical motion, there is no level repulsion in this case. In the
case of chaotic system (three other ensembles in Table
\ref{Ensembles}) the probability to find two levels with close
energies is suppressed. By taking a proper average over the
ensemble of the random Hamiltonian matrices, one can obtain an
energy level distribution satisfactorily for the cluster system
\cite{Stockmann,Dyson,Bohr}. The result is
\begin{equation}\label{ProbN}
P_N(E_1,\ldots,E_N)=C_N^\gamma\text{exp}\left(-\frac{\kappa}{2\delta^2}\sum_jE_j^2\right)\prod_{j<k}\left|E_j-E_k\right|^\gamma,
\end{equation}
where $P_N$ is the probability to find a certain energy level
spectrum, $\gamma=1, 3 \text{ and } 4$ correspond, respectively,
to orthogonal, unitary and sympletic ensemble, $C_N^\gamma$
follows from the normalization condition and $\kappa$ is an
ensemble dependent constant \cite{Dyson}. From the product in
equation \eqref{ProbN} the level repulsion can be clearly seen.

A very important quantity is $R(|E|)$, which can be derived from
the energy level distribution \eqref{ProbN} and gives the
probability to find two energy levels separated by an energy $E$
independent of the number of energy levels in between them
\cite{Dyson}. For the orthogonal ensemble this distribution is
given by
\begin{equation}
\begin{split}
R(|E|)=&1-\left(\frac{\sin\left(\frac{\pi E}{\delta}\right)}{\frac{\pi E}{\delta}}\right)^2\\
&-\frac{d}{d\left(\frac{E}{\delta}\right)}\left(\frac{\sin\left(\frac{\pi E}{\delta}\right)}{\frac{\pi E}{\delta}}\right)\int_{\frac{E}{\delta}}^\infty\frac{\sin(\pi x)}{\pi x}dx. \label{ProbR}
\end{split}
\end{equation}
This expression will be used in the following Section for the
calculation of the transverse dynamic magnetic susceptibility for
a cluster.

Until this point no comments were made about the correctness of
the random matrix theory or the assumption that the energy level
distribution in an irregular cluster is universal and depends only
on the symmetry class. At this point one can say that it is still
a hypotheses that needs to be tested more in order to reach complete
understanding of the situation. However, at the moment there are
many experiments that appear to confirm this theory.
\cite{Stockmann, Halperin, Peerenboom}

\section{Results and discussions}
In this paragraph the Curie temperature of clusters as function of
their size is calculated. From Eq. \eqref{SCRTC} it is clear that
this size dependence could come from $\chi_0$ and $\lambda$. First
the size effect on $\chi_0$ will be considered, second that on
$\lambda$ and finally the resulting effect on the Curie
temperature.

With the use of the probability distribution \eqref{ProbN} it is
possible to calculate the static magnetic susceptibility for a
cluster system. In Ref. 21 an interpolation scheme \mbox{between}
the regimes $T\ll\delta$ and $T\gg\delta$ (or bulk) has been
suggested for which both well developed approximations exist
\cite{Halperin,Denton}. The important result of these calculations
\cite{Peerenboom,Halperin,Denton}, which will be used later on, is
that already for $T>\delta$ (we do not mean strong inequality
here!) the static magnetic susceptibility of a cluster can be
approximated by that of the bulk.

Before $\lambda$ can be calculated, an expression for the
transverse dynamic magnetic susceptibility of a cluster system has
to be found. We will follow analysis originally proposed in Ref.
\onlinecite{Gor} for the case of optical polarizability (further
this result has been slightly corrected, see for review Ref.
\onlinecite{Peerenboom}). The starting point is the general
expression for the transverse dynamic magnetic susceptibility,
\begin{equation}
\chi_0^{-+}(\omega,q)=\sum_\mu\sum_\nu\frac{n(E_\nu)-n(E_\mu)}{E_\mu-E_\nu-\omega+i0}\left|\left\langle\nu\left|e^{i\vec{q}\cdot\vec{r}}\right|\mu\right\rangle\right|^2, \label{SDMX}
\end{equation}
where $|\nu\rangle$ and $|\mu\rangle$ are eigenstates of the
system and $n(E)$ is the Fermi function. Equation \eqref{SDMX}
accounts for a single particle. Therefore, for the Orthogonal
ensemble under consideration, one has to average over the energy
level distribution given in equation \eqref{ProbN} (with
$\gamma=1$). One can then derive that this comes down to
multiplying each term in the sum of equation \eqref{SDMX} by
$\frac{R(|E_\nu-E_\mu|)}{\delta^2}dE_\nu dE_\mu$. For later
convenience the expression for the transverse dynamic magnetic
susceptibility of an Orthogonal ensemble of clusters
$\tilde{\chi}^{-+}_{0}(\omega,q)$ can be written as

\begin{equation}
\begin{split}
\tilde{\chi}^{-+}_{0}(\omega,q)=&\iint dE dE'\iint dE_\mu dE_\nu\delta(E-E_\mu)\delta(E'-E_\nu)\\
&\times\left|\left\langle\nu\left|e^{i\vec{q}\cdot\vec{r}}\right|\mu\right\rangle\right|^2\frac{n(E')-n(E)}{E-E'-\omega+i0}\frac{R(|E'-E|)}{\delta^2}.
\label{iden}
\end{split}
\end{equation}

For $\omega\ll E_F$, which will occur naturally for the
calculation of $\lambda$ at the Curie temperature (Section II), it
was shown in Ref. \onlinecite{Gor} that the matrix element is
approximately energy independent leading to a separation of the
$q$ and $\omega$ dependencies:

\begin{equation}
\begin{split}
&\tilde{\chi}^{-+}_{0}(\omega,q)=A_{\vec{q}}\iint dE dE'\frac{n(E')-n(E)}{E-E'-\omega+i0}\frac{R(|E'-E|)}{\delta^2},\\
&A_{\vec{q}}=\iint dE_\mu dE_\nu\delta(E_F-E_\mu)\delta(E_F-E_\nu)\left|\left\langle\nu\left|e^{i\vec{q}\cdot\vec{r}}\right|\mu\right\rangle\right|^2.
\label{iden2}
\end{split}
\end{equation}

An accurate computation of this $q$-dependence or matrix element
is difficult. However, for $q_c$$\ll$$q$$\ll$$k_F$
(\mbox{$q_c\propto1/R$} is the inverse average radius of the
cluster) the $q$-dependence can be approximated by that of the
bulk, because classical trajectories of electrons in this case are
mainly like in the bulk, with rare reflections at the border.
Thus, within this $q$-regime the real and imaginary part of the
dynamic susceptibility are in \mbox{highest} order proportional
to, respectively $q^2$ and $1/q$ (Eq. \eqref{fbulk}).
\cite{Moriya}

For the $\omega$-dependent part of equation \eqref{iden2} it can be shown that it equals the complex function

\begin{equation}
A(\omega)=2\frac{\omega^2}{\delta}\int\frac{R(|E|)}{E^2-(\omega+i0)^2}dE. \label{AT}
\end{equation}
Besides the approximation of the matrix element, the restriction
of the $q$ regime ($q_c$$\ll$$q$$\ll$$k_F$) has two other
important consequences, one for the evaluation of Eq. \eqref{L1}
and the other for Eq. \eqref{AT}. Here the former will be
discussed first, because the latter will follow from it.

In this regime ($q_c$$\ll$$q$$\ll$$k_F$) the sum over $q$ in Eq.
\eqref{L1} can be replaced by an integral, because the integrand
is a smooth function with a maximum at approximately $q\propto
\omega^{1/3}$, and relevant frequencies are of order of
temperature (here we use units $\hbar=k_B=1$). For the remaining
part of the sum it is assumed that no singularities occur.
Therefore, it is proportional to $1/R$ and can be neglected for
large enough cluster sizes.

\begin{figure}[!h]
  \includegraphics[width=8cm]{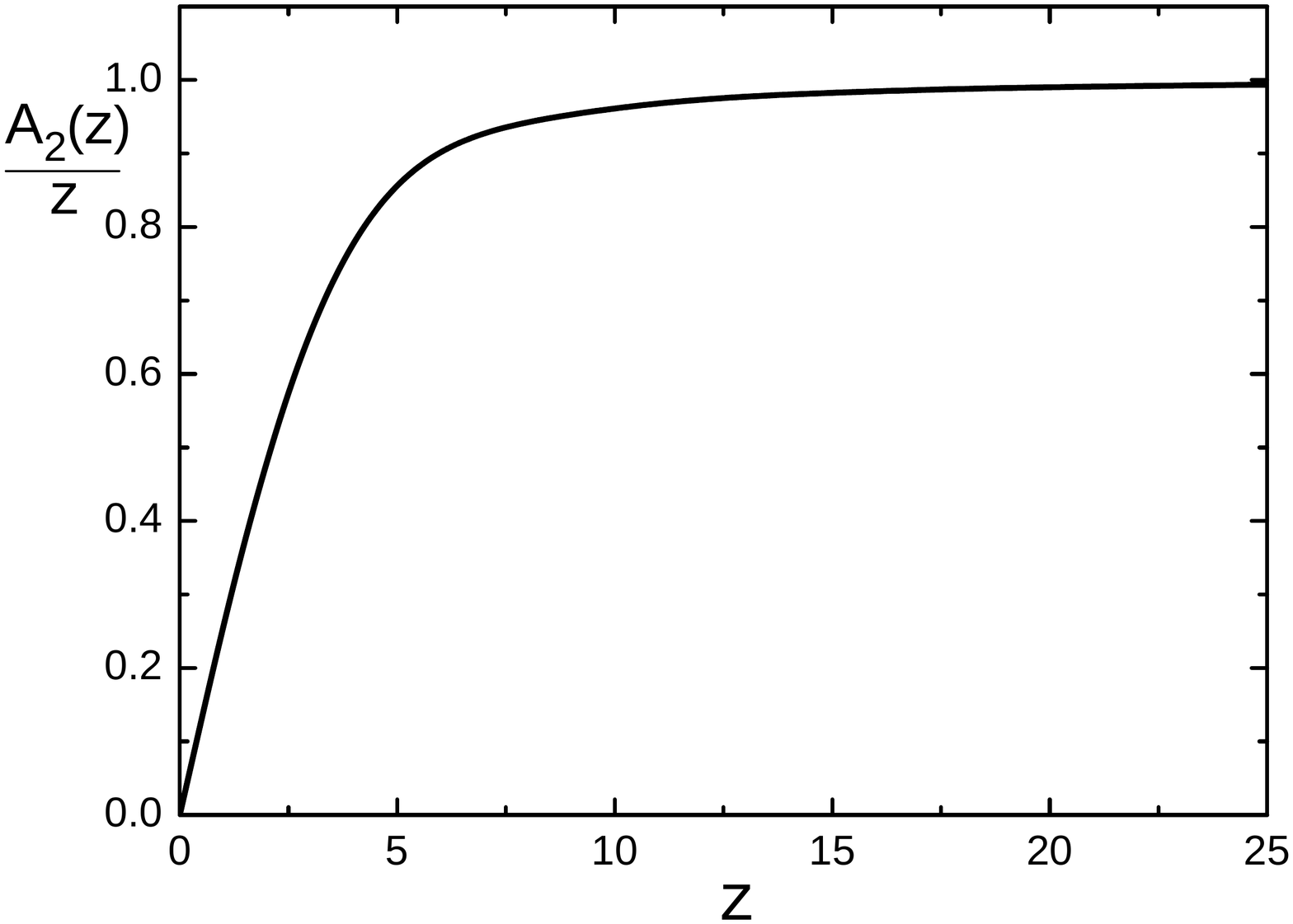}\\
  \caption{The function $A_2/z$ as a function of $z$.}\label{fig4}
\end{figure}

We will thus restrict ourselves only to the case of not too small
clusters, where $\omega\gg\delta$ for relevant $\omega$,
otherwise, the discreteness of $q$-vectors for spin fluctuations
is essential, replacement of sum over $q$ by integral is
impossible, and the problem should be solved numerically for a
given particular shape of the cluster. In this limit one can show
that the $\omega$-dependence of the real part of Eq. \eqref{AT}
can be neglected compared to that of the imaginary part. Then, the
following expression for the transverse dynamic magnetic
susceptibility (normalized to the static susceptibility) of a
cluster system can be obtained
\begin{equation}
f_0(\omega,q)=1-Aq^2+iC\frac{\omega}{q}\frac{A_2(z)}{z}. \label{statf}
\end{equation}
Here $A$ and $C$ are the same constants as for the bulk (equation
\eqref{fbulk}), $z=2\pi\omega/\delta$, the zero in the subscript
of the function $f$ corresponds to $M=0$ (Curie temperature is
found from divergence of susceptibility in paramagnetic regime)
and $A_2$ is the imaginary part of the function $A(\omega)$
\eqref{AT} given by
\begin{equation}
\begin{split}
A_2(z)=&2\eta-\frac{2\sin^2(\eta)}{\eta}\\
&+2\eta\left\lbrack\int_0^\eta\frac{\sin(t)}{t}dt-\frac{\pi}{2}\right\rbrack\frac{d}{d\eta}\left(\frac{\sin(\eta)}{\eta}\right), \label{A2}
\end{split}
\end{equation}
where $\eta=z/2$. At $z$$\rightarrow$$\infty$ $A_2(z)/z$$\rightarrow$$1$ giving extrapolation to the bulk (Figure \ref{fig4}).

At this point it is possible to estimate the cluster sizes for
which the above described approximations are valid. For this
purpose our condition $\omega\gg\delta$ and $\omega\propto T$ can
be used, where the last proportionality can be easily derived from
the calculation of the temperature dependence of $\lambda_1$
(equation \eqref{Lsplit}). Then, using an estimation
$\delta\approx E_F/N$ with $N$ the total number of electrons for
the situation where $E_F=10^4$K and, say, the value $T_c = 20$K
typical for weak itinerant ferromagnets \cite{Moriya} results in
$N>500$. Thus, for five $d$-electrons per atom this would lead to
the condition that the above described considerations are valid
for clusters containing approximately more than 100 atoms.

As was mentioned, in the case $T\gg\delta$ the static magnetic
susceptibility can be approximated by that of the bulk (Eq.
\eqref{NIS}) \cite{Denton}. This means that the important size
dependent contribution to the Curie temperature must come from the
$\lambda_1$ term only. It can be derived by substitution of Eqs.
\eqref{NIS} and \eqref{statf} into Eq. \eqref{L1} and by
approximating the numerators of Eq. \eqref{L1} by their static
long wave-length limit, $\lambda_1$ (here called
$\lambda_{cluster}$), that is,
\begin{widetext}
\begin{equation}
\lambda_{cluster}\propto T^{4/3}\int_0^\infty dx\frac{x^{1/3}}{e^x-1}\left(1-\left\lbrack\frac{\sin\left(\frac{\pi xT}{\delta}\right)}{\frac{\pi xT}{\delta}}\right\rbrack+\left\lbrack\int_0^{\frac{\pi xT}{\delta}}\frac{\sin(t)}{t}dt-\frac{\pi}{2}\right\rbrack\left(\frac{\cos\left(\frac{\pi xT}{\delta}\right)}{\frac{\pi xT}{\delta}}-\frac{\sin\left(\frac{\pi xT}{\delta}\right)}{\left(\frac{\pi xT}{\delta}\right)^2}\right)\right)^{1/3}. \label{Lcluster}
\end{equation}
\end{widetext}
It is justified since for the case of weak itinerant ferromagnets
the region of small $q$ and $\omega$ is dominant in the integral
\eqref{L1}.

\begin{figure}[!h]
  \includegraphics[width=8cm]{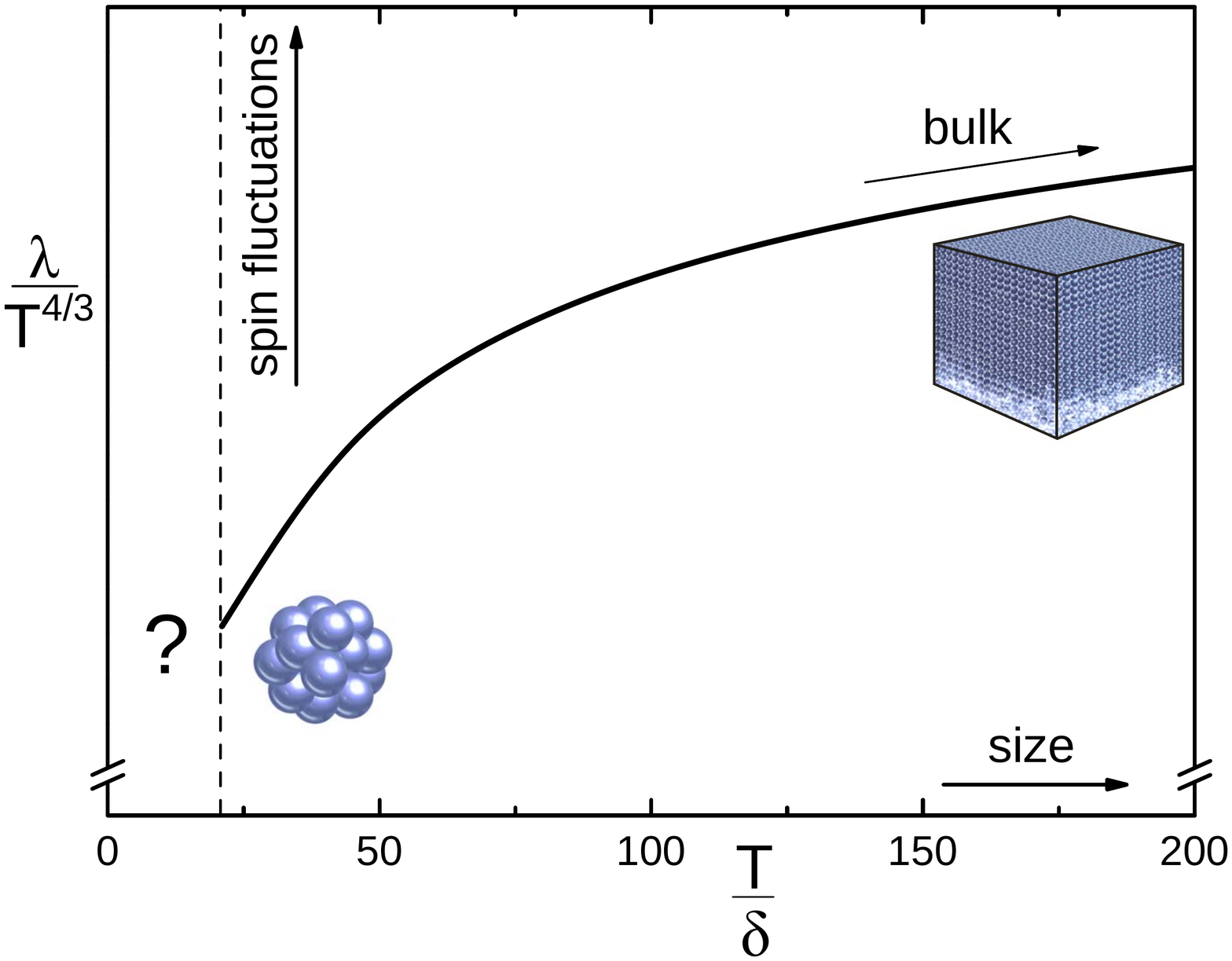}\\
  \caption{(color online) The function $\lambda_{cluster}/T^{4/3}$ as a function of $T/\delta$, indicating the increase of the influence of spin fluctuations at higher temperatures and particulary larger sizes. The question mark indicates the applicability limit of our assumptions.}\label{fig1}
\end{figure}

The result for $\lambda_{cluster}$ is presented as a function of $T/\delta$ in Figure \ref{fig1}.

Obviously, for $\delta$$\rightarrow$$0$ the constant bulk value
\mbox{($\frac{\lambda_{bulk}}{T^{4/3}}=B$)} is reached. Further,
it is important to notice that a cluster system for a fixed size
(or $\delta$) has a larger temperature dependence compared to the
bulk which leads to an increase of the Curie temperature. For
smaller clusters the enhancement of the Curie temperature will be
larger.

\begin{figure}[!h]
  \includegraphics[width=8cm]{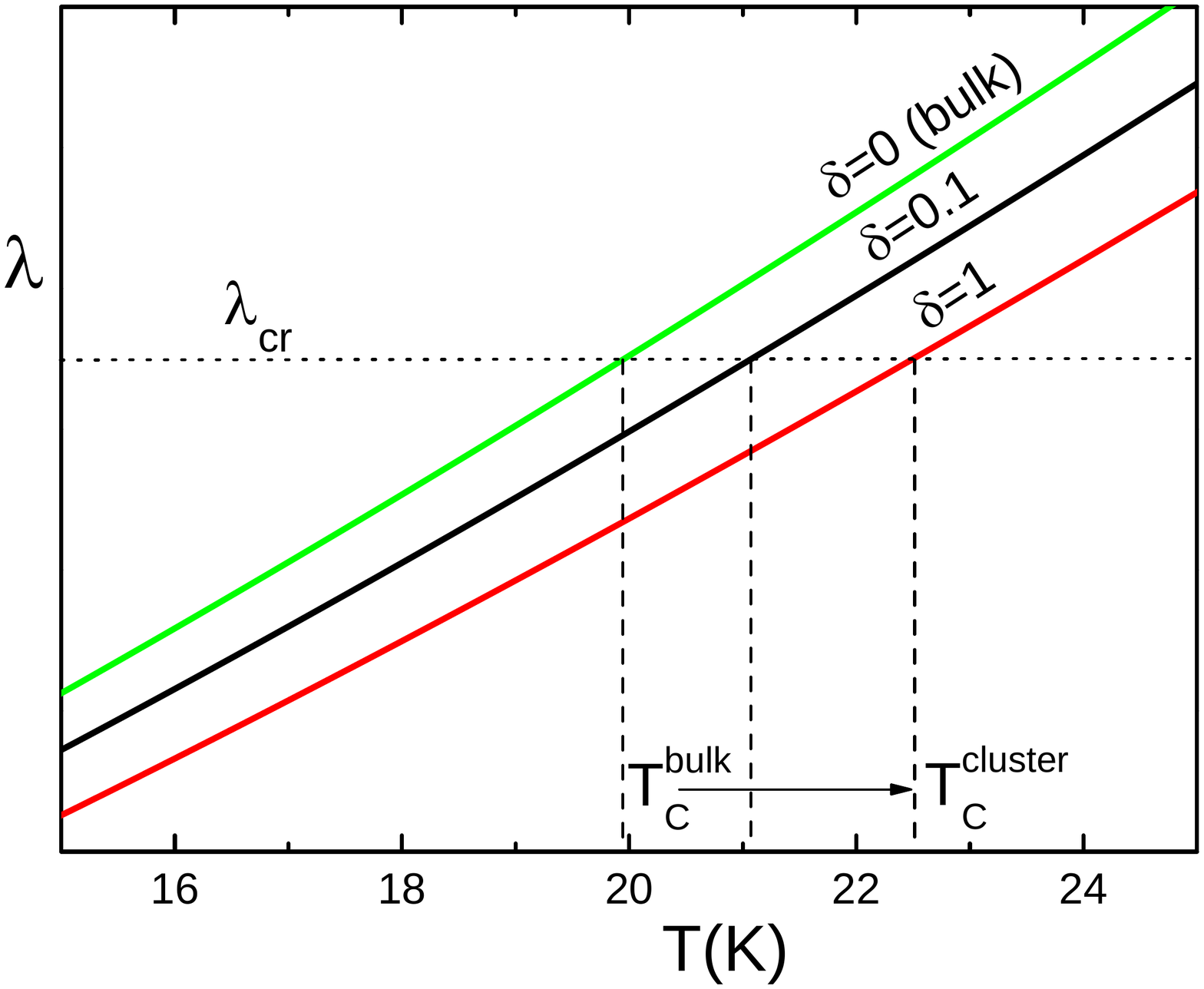}\\
  \caption{(color online) The function $\lambda_{bulk}$ (or $\delta=0$) and $\lambda_{cluster}$ for $\delta=0.1$ and $1$ is plotted as function of temperature. Here $\delta$ is given in units of $K$ and $\lambda_{cr}$ is the critical value of $\lambda$, where the net magnetization becomes zero.}\label{fig2}
\end{figure}

\begin{figure}[!h]
  \includegraphics[width=8cm]{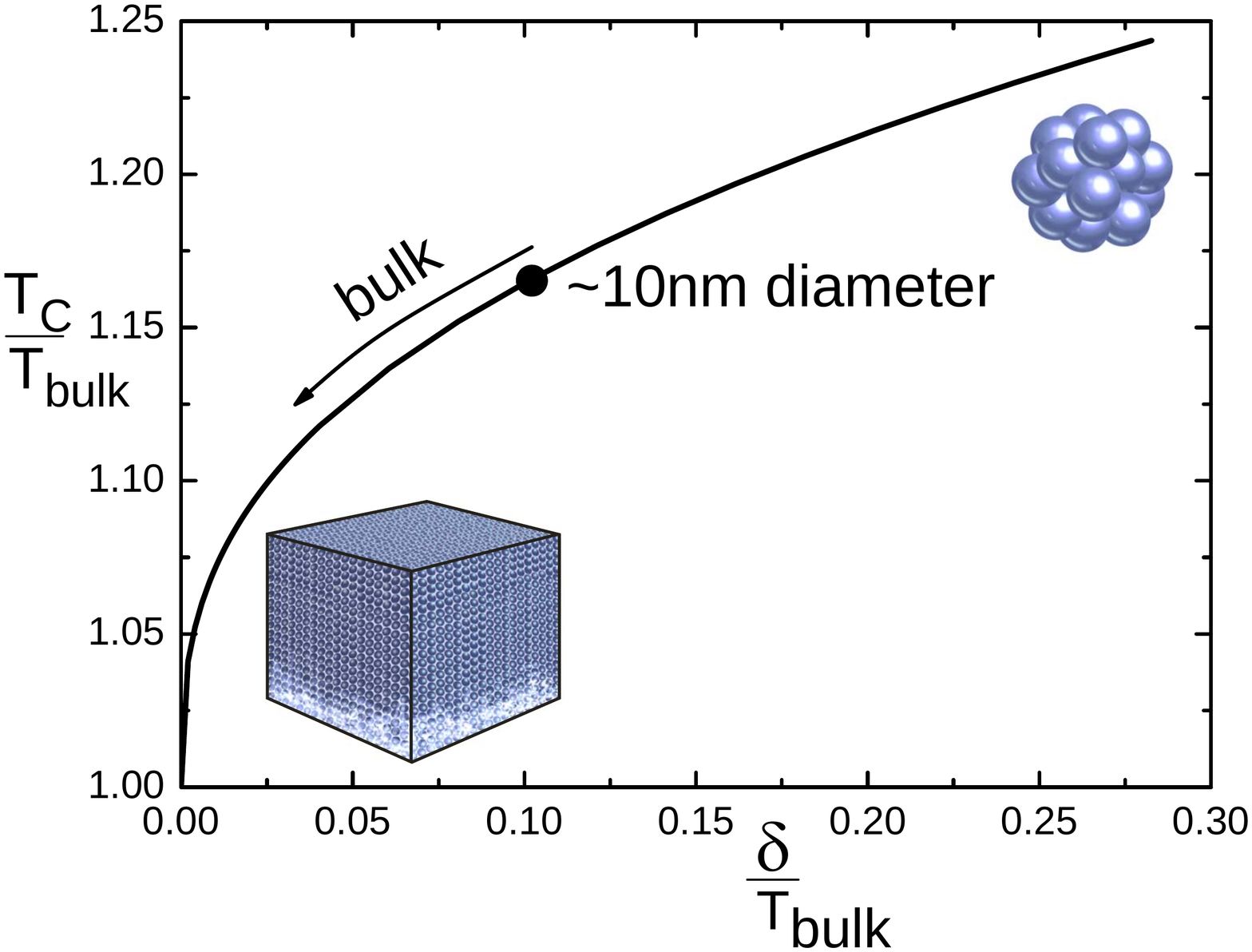}\\
  \caption{(color online) Curie temperature of a cluster system normalized to the Curie temperature of the bulk as a function of the average energy level spacing also normalized to the bulk Curie temperature.}\label{fig3}
\end{figure}

It is instructive to show the data in a slightly different way,
where $\lambda$ is plotted as an function of temperature for
different sizes ($\delta=0$, $0.1$ and $1.0$ in units of $K$) as
can be seen in Figure \ref{fig2}.
To summarize, the results for the Curie temperature of the cluster
normalized to the Curie temperature of the bulk is plotted in
Figure \ref{fig3} as a function of the average energy level
spacing normalized to the Curie temperature of the bulk.
The increase of the Curie temperature for smaller cluster sizes is
caused by the increasing suppression of spin density fluctuations
for smaller sizes. In Section II it was shown that the function
$\lambda$ takes into account the influence of the spin
fluctuations on the Curie temperature.

To verify this theory quantitatively experiments on weakly
itinerant clusters should be performed. We \mbox{believe},
however, that qualitatively our conclusion is correct even beyond
the limit of formal applicability of the SCR theory (equation
\eqref{DefW}). The level repulsion \textit{should} suppress
spectral density of spin fluctuations at small frequencies, and
this \textit{should} be the main effect in the temperature
dependence of magnetic properties, these statements being quite
general. The experiments on Fe, Co and Ni mentioned in the
Introduction \cite{Liu,Bucher,Billas} seem to be in agreement with
our conclusion.

It would be interesting to improve this theory further for smaller
cluster sizes. For this purpose the discreteness of the energy
spectrum should be explicitly taken into account and the influence
of the static susceptibility will become important. For clusters
even smaller so that \mbox{$\delta\propto1/V$} is not applicable
anymore random matrix theory will fail. In this regime probably ab
initio approaches are the only way out.

To conclude, small particles of itinerant magnets show an increase
of their Curie temperature when reducing their size, in a clear
contrast to what one would expect for a localized picture. Such
enhancement of magnetic stability originates from the size induced
renormalization of electronic states leading to a suppression of
spin fluctuations, and may open interesting perspectives for
application of such systems in nanotechnology.

\begin{acknowledgements}
Funding from the European Community's Seventh Framework Programme
(FP7/2007-2013 Grants No. NMP3-SL-2008-214469 (UltraMagnetron) and
No. 214810 (FANTOMAS)] is gratefully acknowledged. \mbox{M. I. K.}
acknowledges a financial support by the EU-India FP-7
collaboration under MONAMI.
\end{acknowledgements}

\end{document}